\documentclass[pdflatex,sn-mathphys-ay]{sn-jnl}

\usepackage{graphicx}%
\usepackage{fontawesome5}
\usepackage{multirow}%
\usepackage{amsmath,amssymb,amsfonts}%
\usepackage{amsthm}%
\usepackage{mathrsfs}%
\usepackage[title]{appendix}%
\usepackage{xcolor}%
\usepackage{textcomp}%
\usepackage{manyfoot}%
\usepackage{booktabs}%
\usepackage{algorithm}%
\usepackage{algorithmicx}%
\usepackage{algpseudocode}%
\usepackage{listings}%
\usepackage{braket}
\usepackage{makecell}
\usepackage{hhline}
\usepackage{tikz}
\usetikzlibrary{positioning, shapes, fit, calc, shadows, backgrounds, arrows.meta}
\usepackage[nolist]{acronym}

\newcommand{\orcidlink}[1]{\href{https://orcid.org/#1}{\color[HTML]{A6CE39}\faOrcid}}

\definecolor{customGreen}{RGB}{130, 193, 157}
\definecolor{customYellow}{RGB}{237, 214, 151}
\definecolor{unitaryGreen}{RGB}{220,236,172}
\definecolor{unitaryBrown}{RGB}{252,204,164}
\definecolor{pathBlue}{RGB}{37,90,124}
\definecolor{predGreen}{RGB}{73,154,127}
\definecolor{predRed}{RGB}{176,36,24}

\theoremstyle{thmstyleone}%
\theoremstyle{thmstyletwo}%

\theoremstyle{thmstylethree}%

\begin{document}
\begin{acronym}
\acro{neb}[NEB]{nudged elastic band}
\acro{aneb}[aNEB]{adapted nudged elastic band}
\acro{qnn}[QNN]{quantum neural network}
\acrodefplural{qnn}[QNNs]{quantum neural networks}
\acro{vqa}[VQA]{variational quantum algorithm}
\acrodefplural{vqa}[VQAs]{variational quantum algorithms}
\acro{qcl}[QCL]{quantum cost landscape}
\acrodefplural{qcl}[QCLs]{quantum cost landscapes}
\acro{qml}[QML]{quantum machine learning}
\acro{vqe}[VQE]{variational quantum eigensolver}
\acro{bp}[BP]{barren plateau}
\acrodefplural{bp}[BPs]{barren plateaus}
\acro{ce}[CE]{concentratable entanglement}
\acro{mlp}[MLP]{multi-layer perceptron}
\acro{hea}[HEA]{hardware-efficient ansatz}
\acrodefplural{hea}[HEAs]{hardware-efficient ansatzes}
\acro{rs}[rs]{resource-scarce}
\acro{ra}[ra]{resource-abundant}
\end{acronym}

\title[Article Title]{Ravines in quantum cost landscapes: opportunities for improved VQA predictions}


\author*[1,2]{\fnm{Felix J.} \sur{Beckmann}\orcidlink{0009-0001-8735-2863}}\email{felix.beckmann@iao.fraunhofer.de}

\author[1]{\fnm{João F.} \sur{Bravo}\orcidlink{0000-0002-8526-5010}}\email{joao.bravo@iao.fraunhofer.de}

\affil[1]{\orgname{Fraunhofer Institute IAO}, \orgaddress{ \city{Stuttgart}, \country{Germany}}}

\affil[2]{\orgname{Karlsruhe Institute of Technology}, \orgaddress{\city{Karlsruhe}, \country{Germany}}}


\abstract{The geometric and topological structure of \acp{qcl} governs the optimization and thus the predictive power of \acp{vqa}. We systematically analyze ravines --- low-cost paths connecting local minima --- using an adapted version of the \ac{neb} algorithm, a method originating from theoretical chemistry. By training \acp{qnn} to classify the concentratable entanglement of quantum states, we apply the \ac{neb} algorithm and numerically identify ravine structures in \acp{qcl} of hardware-efficient ansatzes. Beyond visualizing these ravines, we construct an ensemble prediction framework by averaging predictions from \acp{qnn} parameterized along the low-cost \ac{neb} path.

We introduce a resource-light pre-training metric which quantifies local-prediction variability and serves as a strong performance indicator for \acp{vqa}, even beyond the scope of this study.
When base classifiers are drawn from circuit and weight initializations exhibiting high local-prediction variability, the quantum-based \ac{neb} ensembles outperform both classical and naive quantum alternatives.
Moreover, a complexity analysis shows that leveraging the ravine-like structure of \acp{qcl} with the \ac{qnn} \ac{neb} approach substantially reduces computational costs compared to naive \ac{qnn} ensembling.

A depth and qubit scaling analysis indicates that ravines persist across both scalings, and that, despite the expected growth in resource requirements with the qubit scaling, the \ac{neb} approach also accelerates convergence over the naive alternative.
}


\maketitle

\section{Introduction} \label{sec:introduction}
Classical machine learning is rapidly transforming research and industrial applications, while its quantum counterpart, \ac{qml}, remains in its infancy. Nevertheless, expectations are high: although quantum advantage on classical data is contested, quantum advantage is widely anticipated when \ac{qml} operates on quantum data (\cite{schatzki2021, cerezo2022, chang2025}).

For the successful development of \acfp{vqa}, a thorough understanding of the common properties of cost functions is essential. In particular, the topological and geometrical properties of \acfp{qcl} are of critical importance, both in the NISQ era and beyond (\cite{kathleene.hamilton2021}).

In the classical setting, systematic analysis of the cost landscape has yielded insights that informed both architecture design and optimization (\cite{mackay1992, he2016, keskar2017,  chaudhari2017, li2018}). Research on the nature of the classical cost landscape around the minima proved that most local minima of large decoupled neural networks lie within a band-like structure, with the number of minima outside this band diminishing exponentially with network size (\cite{choromanska2015}). Furthermore, previous research found numerically that the Hessian of the cost landscape decomposes into a near-zero bulk and a small number of outliers (\cite{sagun2018}), indicating the presence of flat directions. Complementing this structural picture, \cite{draxler2019} established flat connectivity between distinct minima in deep networks, concluding that there are essentially no cost barriers in modern deep architectures.

The question of whether \acp{qcl} exhibit comparable structural properties or not has received comparatively little attention, with most research interest in the \ac{qml} community instead concentrated on \acp{bp},  regions of extreme flatness where gradients vanish exponentially with the number of qubits (\cite{jarrodr.mcclean2018, cerezo2021b, larocca2024}). \cite{arrasmith2022} directly linked this phenomenon to narrow gorges, which is the exponential narrowness of minima, proving that both always co-occur.

A particularly relevant contribution to understanding the similarities between classical and quantum cost landscapes comes from \cite{kathleene.hamilton2021}, who examine the connectivity of distinct minima in the QCL for a regression task with classically encoded quantum data in a three-qubit setting. Extending the concept of narrow gorges, they introduce \textit{ravines} --- low-cost connections between minima --- and numerically observe their presence in most of their experiments, while noting that ravines do not imply a globally flat landscape and calling for further investigation. This study follows their definition and defines a ravine as a path between local minima $\mathbf{p}_1$ and $\mathbf{p}_{2}$ for which $\mathcal{C}(\mathbf{p}_r) \lesssim \max \big(\mathcal{C}(\mathbf{p}_1), \mathcal{C}(\mathbf{p}_{2})\big)$ holds, with $\mathcal{C}(.)$ as the cost function and $\mathbf{p}_r$ as any point along the path.

The present study builds upon their analysis of ravines: motivated by the work of \cite{draxler2019} and \cite{kathleene.hamilton2021}, we conduct a systematic analysis of low-cost connections between local minima of \acp{vqa} and extend previous findings to quantum data and to circuits of greater width and depth.
As a further contribution, we systematically explore and visualize the geometric and topological properties of \acp{qcl}, and subsequently introduce a framework that leverages them to improve prediction performance through ensemble learning. Finally, a formal complexity analysis confirms that the presented ensembling approach yields substantial resource savings over naive \ac{qnn} ensembling.
Section~\ref{sec:methods} introduces the key methods, including the \ac{neb} algorithm and the \ac{qnn} filtering metric. The learning setup and \ac{neb} ensembling procedure are detailed in Section~\ref{sec:setup}, followed by the results and their discussion in Section~\ref{sec:results-and-discussion}. The limitations of the proposed framework are discussed in Section~\ref{sec:limitations}, and the paper concludes with an outlook and conclusion in Sections~\ref{sec:outlook} and~\ref{sec:conclusion}, respectively.

\section{Methods} \label{sec:methods}
\subsection{\Acl{neb} algorithm}
\label{sec:neb}
The \ac{neb} algorithm is designed to find a minimum-energy path between two points in a potential landscape. Originally applied to approximate activation energies in theoretical chemistry (\cite{jonsson1998}), the algorithm translates naturally to learning landscapes where the potential corresponds to the cost function $\mathcal{C}$. Specifically, the refined tangent calculation from Henkelman and Jónsson is used, which improves numerical stability at kinks in the path (\cite{henkelman2000}).

To find the low-cost path, a linear interpolation between start and end point is initialized. This interpolation includes $N$ pivot points $\mathbf{p}_i$ whose position will be optimized during the \ac{neb} process by steps in direction of
\begin{align}
    \mathbf{F}(\mathbf{p}_i, k) =  \mathbf{F}^{\mathrm{S}} (\mathbf{p}_{i}, k)\big|_{\parallel} - \mathbf{\nabla} \mathcal{C}(\mathbf{p}_i)\big |_{\perp}.
\end{align}
There are two contributions to the pivot position adjustment:
The first contribution is attractive and acts tangentially to the \ac{neb} path. Thus, it keeps neighboring pivots in proximity as a spring with constant $k$ stretched between neighboring pivots. The force follows Hooke's law as
\begin{align}
    \mathbf{F}^{\mathrm{S}} (\mathbf{p}_{i}, k) = k[(\mathbf{p}_{i+1} - \mathbf{p}_{i}) - (\mathbf{p}_{i} - \mathbf{p}_{i-1})] \cdot \hat{\boldsymbol{\tau}}_{i}\hat{\boldsymbol{\tau}}_{i}
\end{align}
where $\hat{\boldsymbol{\tau}}_i$ denotes the unit tangent to the path at pivot $i$.
The second contribution $\nabla \mathcal{C}(\mathbf{p})\big |_{\perp}$ forces the pivot points in the direction of lower cost values while keeping distance to neighboring pivots by acting only orthogonal to the \ac{neb} path. 

Finally, the pivot position is adjusted with a step size parameter $\eta$ as 
\begin{align}
    \mathbf{p}^{(t+1)} = \mathbf{p}^{(t)} + \eta \, \mathbf{F}.
\end{align}
The resulting pseudo code of the \ac{neb} algorithm is shown in the Appendix~\ref{sec:appendix-alg-neb}.

\cite{draxler2019} extended the NEB algorithm with an outer refinement loop to find low-cost connections between local minima: After each full NEB run, additional pivots are inserted where the linear interpolation of $\mathcal{C}$ surpasses a deviation threshold, evaluated at $(M-1)$ positions between each pair of pivots. At maximum, one new pivot is added between two previous. The NEB is then rerun on this refined pivot set. To prevent the inserted pivots from redistributing uniformly due to the spring force, $k$ is set to 0.

The present work builds on this iterative approach by introducing an adapted step scheduler. The underlying rationale is that newly inserted pivots are initialized between existing ones and thus already reside in the vicinity of a ravine, requiring fewer iterations to converge. This motivates our geometry-aware step scheduler, which progressively reduces the number of steps per \ac{neb} cycle to allow convergence at reduced computational costs. The resulting \ac{aneb} procedure is detailed in Algorithm~\ref{alg:adapted-neb}, with its complexity analysis given in Section~\ref{sec:complexity}.

\begin{algorithm}
\caption{\Acf{aneb} based on \cite{jonsson1998, henkelman2000, draxler2019}}\label{alg:adapted-neb}
\begin{algorithmic}[1]
\Require{Initial path $\mathbf{p} = [\mathbf{p}_0, \mathbf{p}_{1}, \mathbf{p}_{2}, \dots, \mathbf{p}_{N+1}]$ with $N+2$ pivots, a step scheduler $S_{\mathrm{step}}$}
\Function{Adapted NEB}{$\mathbf{p}, S_{\mathrm{step}}$, $\eta$, $k$}
\For{$T$ in $S_{\mathrm{step}}$}
    \State $\mathbf{p} \gets$ \ac{neb}($\mathbf{p}, \eta, T$, $k$)
    \If{not last cycle} 
    \State Calculate $\mathcal{C}$ along $\mathbf{p}$ interpolation
    \State Insert pivots in $\mathbf{p}$ where residuum is large
    \EndIf
    \EndFor
\State \Return $\mathbf{p}^{(T)}$
\EndFunction
\end{algorithmic}
\end{algorithm}

\subsection{Complexity analysis of the \ac{aneb} algorithm}\label{sec:complexity}

The quantum complexity of the \ac{aneb} algorithm (Algorithm~\ref{alg:adapted-neb}) decomposes into two contributions: the complexity of the inner \ac{neb} calls (Algorithm~\ref{alg:neb}), and the overhead incurred between successive \ac{neb} calls. One inner \ac{neb} call together with the subsequent interpolation and pivot insertion step is referred to as one cycle $c \in \{1, \dots, |S_{\mathrm{step}}|\}$, where $S_{\mathrm{step}} = \{T_1, T_2, \dots, T_{|S_{\mathrm{step}}|}\}$ denotes the step scheduler with $T_c$ steps assigned to cycle $c$. Let $p^{(c)}$ denote the number of pivot points at the start of cycle $c$, and let $G$ denote the number of circuit evaluations required to compute the gradient $\nabla \mathcal{C}$ at a single point.

\medskip
\textbf{Inner \ac{neb} complexity}
Each step of the inner \ac{neb} requires evaluating the cost function $\mathcal{C}$ at all $p^{(c)}$ pivots for the tangent estimate, and computing $\nabla \mathcal{C}$ at the $p^{(c)} - 2$ interior pivots. Over $T_c$ steps, cycle $c$ therefore contributes
\begin{align}
    T_c \cdot \left[ p^{(c)} + (p^{(c)}-2) \cdot G \right]
\end{align}
circuit evaluations.

\medskip \textbf{Inter-cycle overhead}
Between successive \ac{neb} calls, the cost function is evaluated along the linear interpolation of the current path to identify segments requiring refinement. Using $M-1$ interpolation points between each pair of neighboring pivots, this amounts to
\begin{align}
    (p^{(c)}-1)M + 1
\end{align}
evaluations per cycle. This overhead is incurred for all cycles except the last.

\medskip \textbf{Combined quantum complexity}
Summing both contributions over all cycles yields the total quantum complexity of the \ac{aneb}:
\begin{align}
    \mathrm{qco_{aNEB}} = \underbrace{\sum_{c=1}^{|S_{\mathrm{step}}|} T_c \cdot \left[ p^{(c)} + (p^{(c)}-2) \cdot G \right]}_{\text{inner NEB calls}} + 
    \underbrace{\sum_{c=1}^{|S_{\mathrm{step}}|-1} \left[(p^{(c)}-1)M + 1\right]}_{\text{inter-cycle overhead}}. \label{eq:adapt-neb-complexity}
\end{align}
Note that eq.~\eqref{eq:adapt-neb-complexity} is an upper bound, since the cost function values computed during the interpolation step could in principle be reused at the first step of the subsequent \ac{neb} call, although this overlap is small and neglected here.

In practice, since $G$ scales linearly with the number of parameters being optimized, the gradient evaluations dominate the complexity asymptotically and $\mathrm{qco_{aNEB}}$ is well-approximated by retaining only the $G$-weighted terms.

\medskip \textbf{Naive-emsemble complexity}
For comparison, the naive approach trains $k \in \mathbb{N}$ independent models for $T \in \mathbb{N}$ gradient steps each, giving
\begin{align}
    \mathrm{qco_{naive}} = k \cdot T \cdot G. \label{eq:naive-complexity}
\end{align}
Optimizer-specific contributions are typically negligible relative to the gradient evaluations and thus are omitted.

\medskip \textbf{Comparison}
A key advantage of the \ac{neb} approach is that pivot points can be inserted progressively across cycles, as a ravine-like \ac{qcl} structure ensures that even sparsely initialized \ac{neb} settings converge to meaningful low-cost connections. Therefore, ensemble members along a converged path yield effective predictions despite being added in later \ac{neb} cycles. Consequently, the initial pivot count can be kept small, with refinements added only where needed, reducing the dominant $G$-weighted gradient evaluations in the early cycles.

\subsection{Ensemble learning}
\label{sec:ensemble-learning}
Ensemble learning is a well-established method for improving the prediction capabilities of classifiers. Rather than relying on a single classifier, a set of distinct classifiers is trained, and their predictions are combined into a joint ensemble prediction. The combination can be done in various ways, with averaging over individual predictions being a common approach (\cite{zhou2012}).

The performance of an ensemble classifier can be characterized by two quantities: the prediction \textit{accuracy} of the individual classifiers and the degree of \textit{independence} between them (\cite{breiman2001}). This reflects the intuition that a good ensemble consists of individually capable classifiers that nevertheless make sufficiently distinct errors (\cite{krogh1994}).

\subsection{Prediction-correlation metric}
\label{sec:prediction-correlation-metric}
Ensemble performance depends on selecting base classifiers from favorable regions of a \ac{qcl}. However, the choice of a suitable quantum circuit architecture and weight initialization presents a significant challenge due to the complex structure of \acp{qcl} (\cite{ge2022, sonnyrappaport2023, mhiri2025}).
To address this, we introduce a landscape quality metric that enables systematic evaluation of circuit and initialization choices \textit{prior} to any optimization procedure. The metric builds upon the relative fluctuation $\sigma$ introduced by \cite{zhang2024}, who proposed it as an optimization-free indicator of \ac{qcl} learnability in a \ac{vqe} setting. It combines the variance of the energy expectation value over the parameter space $\Theta$ with a normalization factor to yield an inter-circuit comparable measure:
\begin{align}
    \sigma \propto \sqrt{\underset{\theta \in \Theta}{\mathrm{Var}}(\langle H\rangle)}.
\end{align}
Zhang et al. demonstrate that this global metric captures key structural properties of \acp{qcl}, including \acp{bp}, weak expressibility, and ruggedness, which is corroborated through numerical simulations. The present work builds upon this metric and adapts it to the typical \ac{qml} setting of multi-state classification with a \ac{vqa}.
To generalize the metric to this setting, the \textit{correlation} between prediction vectors is employed as an effective measure of similarity.

A key distinction of the metric introduced here is its \textit{locality}. Since gradient-based optimizers with a finite learning rate explore only a bounded neighborhood of the initialization point, the accessible region of parameter space is constrained by the product of the learning rate, the number of optimization steps, and the magnitude of the gradient, rendering distant regions of the landscape irrelevant to convergence behavior. Consequently, the structure of a \ac{qcl} in the vicinity of the initialization is most consequential for optimization success, whereas a metric defined over the full parameter space $\Theta$ may not faithfully reflect the learnability experienced in practice. The proposed metric therefore samples prediction vectors within a Gaussian-weighted neighborhood of a reference point in parameter space, effectively localizing the evaluation to the region relevant for training. Slow, large-scale variations in the prediction landscape are thus appropriately down-weighted.

The normalization factors from Zhang et al. are omitted here: the restriction to a single measured Pauli operator renders the Hamiltonian-norm factor redundant, while the Pearson correlation coefficient intrinsically normalizes by the sample variance, resulting in a direct measure of similarity.
Combining these considerations yields the landscape quality metric
\begin{align}
    \zeta(\theta;s) = \underset{\theta_1, \theta_2 \sim \mathcal{N}(\theta, s) }{\mathbb{E}} \Big[ R\big(\hat{y}(\theta_1),\hat{y}(\theta_2) \big) \Big],
\end{align}
where $R(\cdot,\cdot)$ denotes the Pearson product-moment correlation coefficient between two prediction vectors $\hat{y}$, each obtained for a given $\theta$. The applied prediction function  $\hat{y}$ is defined in Equation~\ref{eq:y-hat}. The scale parameter $s$ controls the width of the Gaussian-weighted neighborhood $\mathcal{N}(\theta, s)$, which should be chosen to reflect the expected parameter displacement per optimization step. A low value of $\zeta$ indicates high variability in the predictions within the local neighborhood, corresponding to a circuit and initialization that is suitable for optimization.

A key practical advantage of this metric is that it can be evaluated \textit{prior} to training, at the stage where circuit architecture and weight initialization are selected. Therefore, one may repeat the sampling procedure until $\zeta$ falls below a chosen threshold. Since no gradient evaluations are required, the computational overhead of this procedure is negligible relative to the subsequent gradient-based optimization.

\section{Experimental setup} \label{sec:setup}
\subsection{Learning problem}
\begin{figure}
    \includegraphics[width=\linewidth]{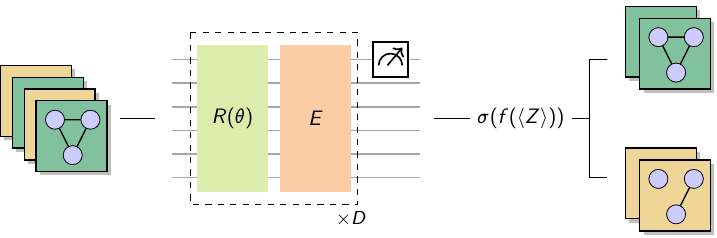}
    \caption{Schematic overview of the classification pipeline. Input quantum states from two classes of distinct \ac{ce} are processed by a \ac{qnn} comprising $D$ alternating layers of parameterized rotational gates $R(\boldsymbol{\theta})$ and entangling gates $E$. The Pauli-$Z$ expectation value of one of the first three qubits is measured, linearly rescaled by $f$, and passed through a Sigmoid activation $\sigma$ to produce a class prediction. States are assigned to the low- or high-entanglement class based on a decision threshold of $0.5$.}
    \label{fig:ntangled-classification}
\end{figure}

A controlled comparison across network architectures and optimization procedures requires a well-defined learning problem comprising a dataset, a prediction network, and a cost function, alongside the optimization procedures to be compared. The resulting classification process is depicted in Figure~\ref{fig:ntangled-classification}.

\medskip \textbf{Datasets} This study requires data sets that scale naturally with the number of qubits and offer tuneable difficulty are desirable for a meaningful comparison. The NTangled dataset satisfies both requirements: it provides a method to generate quantum states of various sizes whose \ac{ce} is distributed around a specified target value (\cite{schatzki2021}). The \ac{ce} of an $n$-qubit state $\ket{\Psi}$ is defined as
\begin{align}
    \mathrm{CE}(\ket{\Psi}) = 1-\frac{1}{2^n} \sum_{\alpha \in Q} \mathrm{Tr}[\rho_{\alpha}^2], \label{eq:ce}
\end{align}
where $Q$ denotes the power set of $[n]$ and $\rho_{\alpha}$ is the reduced density operator of subsystem $\alpha$. Using quantum data natively avoids the encoding overhead that has been identified as a source of trainability challenges in several studies (\cite{bowles2024, larocca2024}).

The learning problem is formulated as a binary classification between two classes of quantum states with distinct \ac{ce} values. Two datasets are considered: one with $\mathrm{CE} \in \{0.05, 0.35\}$ and one with $\mathrm{CE} \in \{0.15, 0.35\}$. We assume the latter, with the smaller separation between target \ac{ce} values, is a harder classification task since the entanglement distributions of the states show a bigger overlap. The hyperparameters for the generation of input states, listed in Table~\ref{tab:ds-parameters}, were chosen to minimize inter-class overlap in the \ac{ce} distributions. Nevertheless, the labeling procedure introduces an expected false-label rate of approximately $5\%$, which sets an upper bound of $\approx95\%$ on the average accuracy achievable on unseen data.

Since estimating the \ac{ce} requires access to all subsystem purities, multiple copies of the input state are necessary (\cite{cincio2018, beckey2021}). Throughout this study, two copies are provided, meaning that 3-qubit states are generated and concatenated into a single 6-qubit input, except where noted i.e. in the scaling experiments.

Finally, each dataset is split into 400 training and 600 test states, allowing for a precise evaluation on test data. To reduce dataset dependency, three data shuffle seeds are used and their results are aggregated in the evaluation.

\medskip \textbf{Cost function} To evaluate the predictions made by the networks, a binary cross entropy cost $\mathcal{C}_{\mathrm{BCE}}$ is used.

\begin{align}
    \mathcal{C}_{\mathrm{BCE}} = -\left( y \log(\hat{y}) + (1 - y)\log(1 - \hat{y}) \right)
\end{align}

\subsection{Network architectures}

\medskip \textbf{\Aclp{qnn}} This study is oriented towards generalization across learning problems, requiring circuits that are both resource efficient and versatile. \Acp{hea} satisfy both criteria through their generic structure and low circuit depth (\cite{cerezo2021a}).

\begin{figure}
    \centering
    \includegraphics[width=1\linewidth]{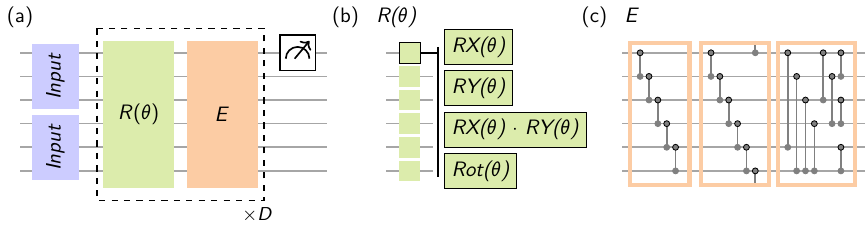}
    \caption{The structure of the \ac{qnn} architecture:
    \textbf{(a)}~Overall circuit layout, consisting of at least two copies of the input state, followed by $D$ alternating rotational and entangling layers, with the Pauli-$Z$ expectation value measured on one of the first three qubits.
    \textbf{(b)}~The rotational layer, consisting of a set of single-qubit gates shared across every qubit. For each circuit, gates are drawn from parameterized Pauli-$X$, Pauli-$Y$, a combination of both, or arbitrary rotation gates with rotation parameter $\theta$.
    \textbf{(c)}~The entangling layer $E$, realized through one of three topologies, each using CNOT gates with the encircled qubit as control: cyclic, chain, and all-to-all connectivity.}
    \label{fig:circuit-layout}
\end{figure}

Circuits are sampled to cover a broad range of architectural configurations. Each circuit consists of three parts: all circuits start by loading the copies of a quantum state onto the registers, followed by alternating rotational and entangling layers. Rotational layers draw one kind of parameterized gates (Pauli-$X$, Pauli-$Y$, their combination, or arbitrary rotation gates) while entangling layers adopt chain, cyclic (\cite{kathleene.hamilton2021}), or all-to-all connected topology. Figure~\ref{fig:circuit-layout} illustrates the resulting layout. Excluding the scaling analysis, six circuits are evaluated, with depths ranging from 2 to 10 and parameter counts between 12 and 72.

The output is obtained by measuring the Pauli-$Z$ expectation value on one of the first three qubits. This value is linearly rescaled with factor $5$ and bias $b$ and passed through a sigmoid function to improve gradient flow across the observable's value range, yielding the prediction
\begin{align}
    \hat{y} = \sigma\!\big(5 \cdot \langle Z \rangle_{\theta} + b\big)
    = \frac{1}{1 + \exp\!\Big(-\big(5 \cdot \langle Z \rangle_{\theta} + b\big)\Big)}. \label{eq:y-hat}
\end{align}
The bias parameter $b$ is optimized along the rotational parameters of the \acp{qnn}.

All \acp{qnn} are simulated using PennyLane (\cite{pennylane2022}) with JAX (\cite{jax2025github}) for computational acceleration, in a noise-free setting throughout.

\medskip \textbf{\Acl{mlp}} As a classical counterpart to the \ac{hea}, a \ac{mlp} was included to enable a fair classical comparison. To match the parameter counts of the sampled circuits, architectures with either one or two hidden layers were used, covering a range of 19 to 109 parameters. The input vector was constructed by separating the complex amplitudes of the quantum states into their real and imaginary parts. ReLU activations were applied in the hidden layers and a Sigmoid activation on the output. As overfitting was observed during preliminary experiments, the training dataset of the \acp{mlp} is further partitioned into training and validation subsets at a ratio of 4:1, with early stopping conducted with the validation data results.

\subsection{Training protocols}
\label{sec:training-protocols}
\medskip \textbf{\ac{neb} optimization protocol}
Two minima were first obtained via standard Adam optimization and subsequently used as initialization points for the adapted NEB algorithm (Algorithm~\ref{alg:adapted-neb}).

Two hyperparameter configurations were employed, each targeting a distinct objective. The first configuration aimed to confirm the \textit{existence} of a low-cost connection in the cost landscape, and was therefore allocated sufficient resources to converge to such a path, hence its label as the \textit{resource-abundant} configuration. It uses a large number of cycles and iterations, inserts additional pivots wherever the deviation exceeds a predefined threshold, and sets the spring constant to zero. The latter ensures that pivots are not distributed uniformly along the path. Instead, all pivots and their interpolations should aim at the cost level of the initialization points.

The second configuration addresses whether low-cost paths are well-suited for \textit{practical ensemble prediction}. Here, the emphasis shifts from convergence guarantees to computational efficiency: the goal is to identify sufficiently good points along the low-cost connection without excessive resource expenditure, hence its label as \textit{resource-scarce} configuration. The exact parameter choices for both configurations are listed in Tables~\ref{tab:neb-parameters} and \ref{tab:scheduler-parameters}.

In both settings, the step scheduler is designed such that the first cycle comprises more iterations, providing a rough approximation of the potential ravine, which is then refined in subsequent cycles.

\begin{figure}
    \centering
    \includegraphics[width=\linewidth]{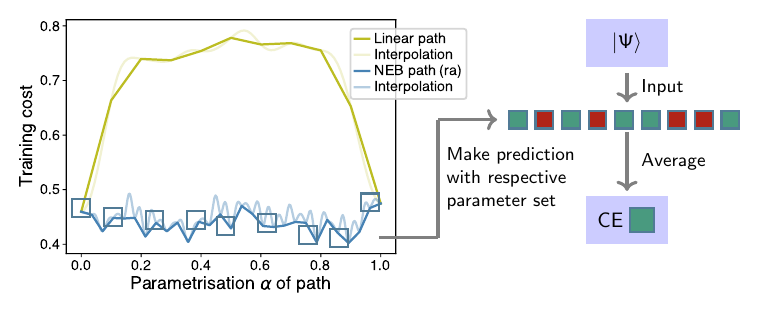}
    \caption{Illustration of the introduced \ac{neb} ensemble prediction framework. Pivot points along the \ac{neb} path each define a classifier configuration. A subset of these classifiers is selected and their predictions of the \acf{ce} class for input states $\ket{\Psi}$ are aggregated across the ensemble, here via averaging. Here, the \acf{ra} \ac{neb} configuration was applied.}
    \label{fig:neb-ensemble-framework}
\end{figure}

Ensemble predictions are obtained by evaluating the predictions of a set of pivot points along a low-cost path, each corresponding to a distinct classifier configuration, and averaging their predictions per data point, as illustrated in Figure~\ref{fig:neb-ensemble-framework}.

\medskip \textbf{Naive ensemble}
In the naive ensemble approach, each base classifier is trained separately, without coordination. At inference time, predictions are aggregated by averaging across ensemble members. To ensure a fair comparison, the number of independently trained classifiers equals the total number of pivots in the final \ac{neb} path, and both approaches are allocated the same number of iterations using Adam as the optimizer which in the \ac{aneb} case are only used by pivots added in the first cycle. Consequently, the training of one naive ensemble classifier is identical to one initialization point acquisition of the \ac{neb} procedure.

\subsection{Experimental baselines}
This work investigates the potential of the \ac{neb} procedure for ensemble prediction. To isolate the sources of any observed advantages, two axes of comparison are considered: whether \acp{qnn} outperform \acp{mlp} on the learning problem, and whether \ac{neb} ensembling outperforms the naive ensemble approach. All four combinations are evaluated.

For the \ac{qnn} results, the prediction-correlation metric $\zeta$ introduced in Section~\ref{sec:prediction-correlation-metric} is used as a pre-training filter. Since $\zeta$ can be evaluated before training with negligible computational overhead, it provides a principled criterion for selecting promising circuit and initialization combinations. Results are reported for different percentile thresholds of $\zeta$, retaining only the configurations with the lowest values.

Since $\zeta$ is specifically designed for \acp{qnn}, \acp{mlp} are instead filtered by a dedicated metric $\chi$, defined as the minimum validation cost averaged across the classifiers of one ensemble. Ensembles are ranked by this quantity, and only those falling within the lowest $x$-th percentile are retained for analysis.

Crucially, this \ac{mlp} filter $\chi$ is computed post-training and directly reflects optimization outcomes, whereas the \ac{qnn} filter is computed prior to training. Any performance advantage of \acp{qnn} over \acp{mlp} in the filtered results can therefore be attributed to model quality rather than a more favorable filtering procedure.

\subsection{Complexity analysis}
\label{sec:absolute-complexity}
The presented framework is a quantum-classical hybrid algorithm. Given that the wall-clock time of operations on quantum hardware is significantly bigger than for operations on classical hardware, the complexity analysis focuses on the quantum complexity of the algorithms. The derivation for the \ac{aneb} algorithm (Algorithm~\ref{alg:adapted-neb}) is provided in Section~\ref{sec:complexity}. Following the parameter shift rule~(\cite{mariaschuld2019}), the cost of a single gradient evaluation is assumed to be $2d$, where $d = \dim(\nabla \mathcal{C})$.

The quantum complexity of the entire \ac{neb} workflow includes the complexity of the initialization point generation and the resources for the actual \ac{aneb} procedure.
Thus, the quantum complexity of the \acf{rs} \ac{neb} configuration evaluates to
\begin{align}
    \mathrm{qco}_{\mathrm{NEB}} &= \mathrm{qco}_{\mathrm{init}} + \mathrm{qco}_{\mathrm{aNEB}} \big|_{\mathrm{rs}} \\
    &= 2000d + (4600d + 3337) \\
    &= 6600d + 3337,
\end{align}
where $d$ denotes the dimensionality of the search space. For comparison, the quantum complexity of naive ensemble training is derived in Section~\ref{sec:complexity} (Equation~\ref{eq:naive-complexity}). The comparable case consists of $9$ independently trained models with $500$ steps each, yielding
\begin{align}
    \mathrm{qco}_{\mathrm{naive}} = 9000d.
\end{align}

In the high-$d$ regime, the relative quantum complexity savings of the \ac{neb} workflow approach $\approx 27\%$ compared to naive ensemble training. This advantage arises because pivots are added incrementally across cycles: pivots inserted in later cycles only the resources required for the remaining cycles, rather than the full procedure, reducing the total number of gradient evaluations compared to running the full optimization for all classifiers as in the naive ensembling.

\subsection{Scaling experiments}
\label{sec:experimental-details-scaling}

Since this study relies on classical simulations of quantum circuits, computational constraints limit the tractable circuit sizes. Nevertheless, understanding how the proposed methods scale with circuit width and depth is essential for assessing their viability on future noise-resilient hardware. Two questions are addressed: whether low-cost connections are still found for larger circuits, and how resource requirements scale with the search space-dimensionality and the number of qubits.

Due to computational restrictions, the scaling analysis is limited to the two best-performing circuits from the main experiments and a single data-shuffle seed. Depth scaling is examined by doubling and quadrupling the number of layers, while width scaling extends the analysis to 8- and 9-qubit circuits. For the width scaling experiments, the number of \ac{neb} steps is tripled to allow for a more precise scaling analysis given the increased resource demands.

Changing the circuit width requires adapting both the dataset and the circuit architecture accordingly. For 8-qubit circuits, a duplicated 4-qubit input state is used. The generation follows the procedure from \cite{schatzki2021} with the parameters provided in Table~\ref{tab:4q-ds-parameters} For 9-qubit circuits, the 3-qubit states from the main experiments are triplicated.

\begin{figure}
    \centering
    \includegraphics[width=\linewidth]{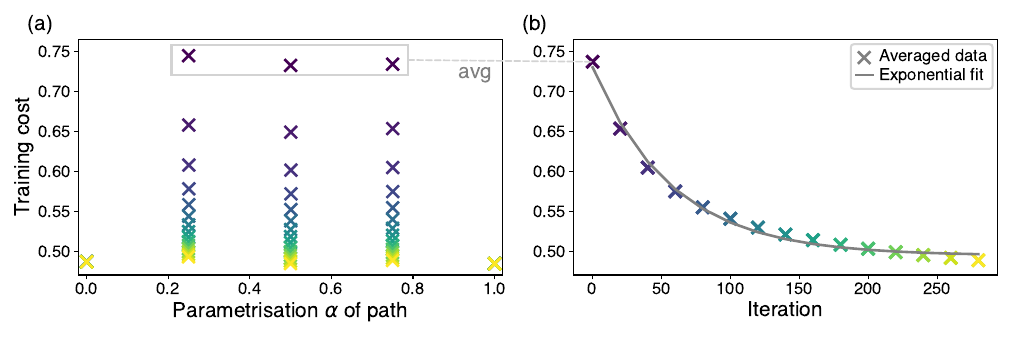}
    \caption{Schematic overview of the convergence analysis applied in the width scaling experiments. Subfigure~\textbf{(a)} shows the training cost trace of a \ac{neb} optimization, with the iteration encoded in color. For each iteration, the cost values of the newly added pivots are averaged to yield a single data point, shown in Subfigure~\textbf{(b)} alongside an exponential fit of the form $f(x) = ae^{bx}+c$, whose decay parameter $b$ serves as the relevant quantity for comparing convergence speed.}
    \label{fig:width_decay_explanation}
\end{figure}

Figure~\ref{fig:width_decay_explanation} illustrates the procedure of analyzing the convergence behavior for different widths: the gathered training cost values are averaged over the pivots added in the respective cycle (Fig.~\ref{fig:width_decay_explanation}(a)). The resulting values are then used to conduct an exponential fit of the form $f(x) = a\cdot\exp(bx) + c$ with parameters $a,b,c \in \mathbb{R}$. This exponential model is motivated by the expected exponential convergence of gradient descent in the vicinity of a local minimum. It is further adopted for its simplicity, as it summarizes the decay rate in a single interpretable parameter $b$, which can be directly compared across different settings. Since the point of convergence of all pivots changes each cycle with the addition of new pivots, each cycle can be regarded as an independent decay process. As the most relevant process is the decay into the ravine corresponding to the first cycle in which a pivot was added, this decay analysis is restricted to just the first cycle of each pivot. Given the decay parameter $b$, the number of iterations required to reduce the cost value to a threshold of $\delta=5\%$ of its starting value relative to the point of convergence can be computed as
\begin{align}
    \mathrm{it}^{\star} = \frac{\log\delta}{b}.
\end{align}

Although scaling experiments are limited in scope, they provide indicative evidence of whether the proposed framework remains viable beyond the primary experimental setting.

\section{Results \& discussion} \label{sec:results-and-discussion}
\subsection{Existence of ravines}

Previous research suggested the existence of ravines in \acp{qcl} in a 3-qubit setting when learning from classical data (\cite{kathleene.hamilton2021}). We extend this result and confirm the existence of such ravines across the vast majority of tested six qubit settings. All cost values reported in this subsection are evaluated on the training set, using the resource-abundant \ac{neb} configuration.

\begin{figure}
    \centering
    \includegraphics[width=\linewidth]{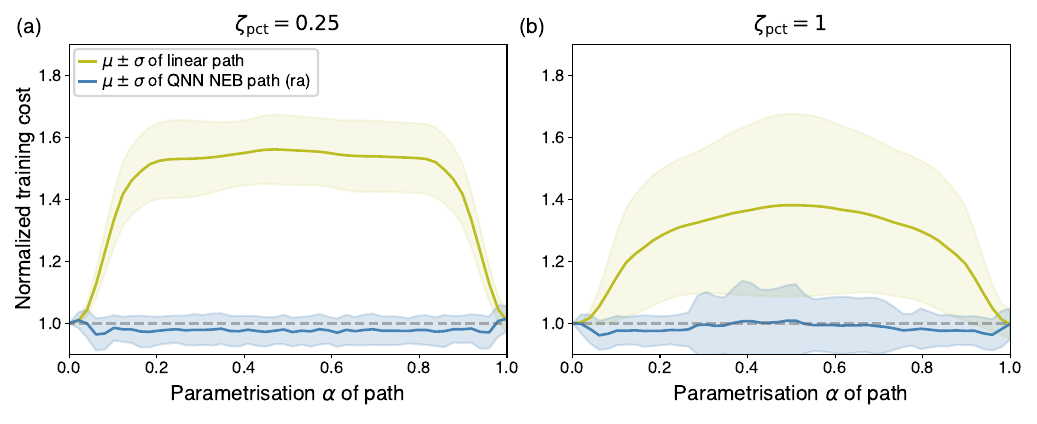}
    \caption{Both Subfigures show the \ac{qnn} training cost normalized to its value at $\alpha=0$, evaluated along both a linear interpolation and an optimized \ac{neb} path between the initialization points gathered in the \acf{ra} \ac{neb} configuration. Subfigures~\textbf{(a)} and~\textbf{(b)} correspond to filtering by the $25$th and $100$th percentile of the prediction-correlation metric, respectively, computed prior to optimization. The solid lines represent the average across configurations, while the shaded areas indicate the corresponding $\pm 1\sigma$ band.}
    \label{fig:average_path_combined}
\end{figure}

Figure~\ref{fig:average_path_combined} shows the normalized cost values along the optimized \ac{neb} path. For low values of the prediction-correlation metric ($\zeta_\text{pct} = 0.25$), a clear distinction between the linear interpolation and the optimized path is apparent, corresponding to the finding of ravines. While the cost values along the linear connection form a plateau with steep ends, the average \ac{neb} training cost remains below that of the initialization points at every point along the optimized \ac{neb} path. This suggests not only the existence of ravines, but also that classifiers located within the ravine achieve better training performance than the initialization points, which were naively trained with comparable resources. One possible explanation is that, while ravines are clearly distinguishable from plateaus by their lower cost values, they are not entirely flat internally. Pivots already residing within a ravine continue to be optimized, ultimately converging to regions of even lower cost than those reached by the naive initialization points.

When aggregating over all circuit architectures and initializations ($\zeta_\text{pct} = 1$), this trend of finding ravines persists but is attenuated: the linear connection forms a less pronounced plateau with considerably higher variance, and the optimized \ac{neb} path likewise exhibits higher variance, with cost values that do not consistently fall below those of the initialization points. Inspection of individual traces reveals isolated settings in which no low-cost connection could be identified by the \ac{neb} procedure, as well as cases in which the linear interpolation is already flat at initialization.

For classical neural networks, low-cost connections between local minima are well established (\cite{ballard2016, ballard2017, freeman2017, draxler2019}). \cite{draxler2019} find consistently that the barriers along such connecting paths vanish --- turning them into proper ravines --- in wider and especially deeper convolutional neural networks, hinting at a positive link between number of parameters and ravines in classical loss landscapes.
Nevertheless, ravines in \acp{qcl} have only been observed in small likely-underparameterized \ac{qnn} circuits. Thus, it is important, and likely useful, to analyze whether ravines arise in larger overparameterized \ac{qnn} circuits. Overparameterization necessarily leads to redundant dimensions in the loss landscape, which can embed local minima points into higher-dimension spaces with similar loss value, and thus may aid the formation of ravines.

\begin{figure}
    \centering
    \includegraphics[width=1\linewidth]{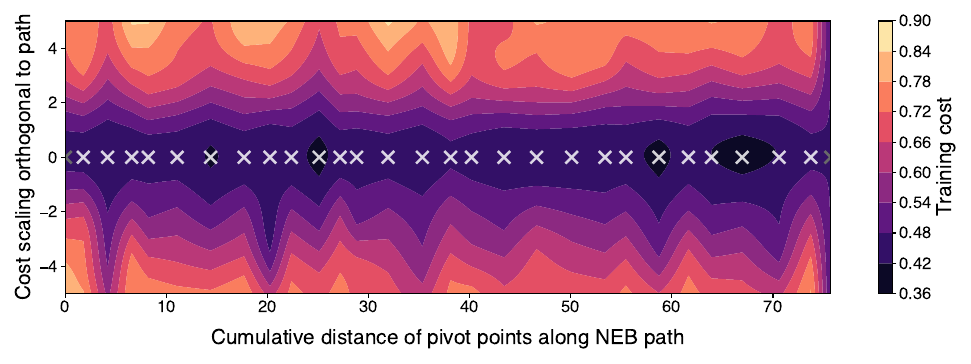}
    \caption{Visualization of a \ac{qnn} \ac{neb} path between two initialization points. The path is straightened along the $x$-axis, with the $y$-axis representing the direction orthogonal to it. The orthogonal component at pivot $i$, $\mathbf{p}_i$, is calculated in the plane spanned by the tangent at $\mathbf{p}_i$ and the connection of pivot $(i-1)$ and $i$, $\overline{\mathbf{p}_{i-1} \mathbf{p}_{i}}$. The color encodes the training cost $\mathcal{C}_{\mathrm{tr}}$, and the gray and white crosses mark the initialization and pivot points, respectively.}
    \label{fig:single_neb_path}
\end{figure}

Figure~\ref{fig:single_neb_path} provides an intuitive illustration of the ravine structure of a \ac{qcl} for a single exemplary optimized \ac{neb} path. The path is unfolded along the horizontal axis, with the cumulated distance between pivot points used as the parameterization, while the vertical axis encodes displacements orthogonal to the \ac{neb} path. For most pivot points, the cost values are lowest along the path itself and increase rapidly away from it, confirming the narrow, channel-like geometry characteristic of a ravine. Nevertheless, some pivot points exhibit a broader low-cost neighborhood, suggesting the presence of side tracks or branching ravines merging into the optimised \ac{neb} path.

\subsection{Ensemble performance}
After confirming the existence of ravines, this section examines the properties of the pivot points along the \ac{neb} path optimized in the resource-scarce setting, with all reported metrics evaluated on the held-out test set.

\begin{figure}
    \centering
    \includegraphics[width=1\linewidth]{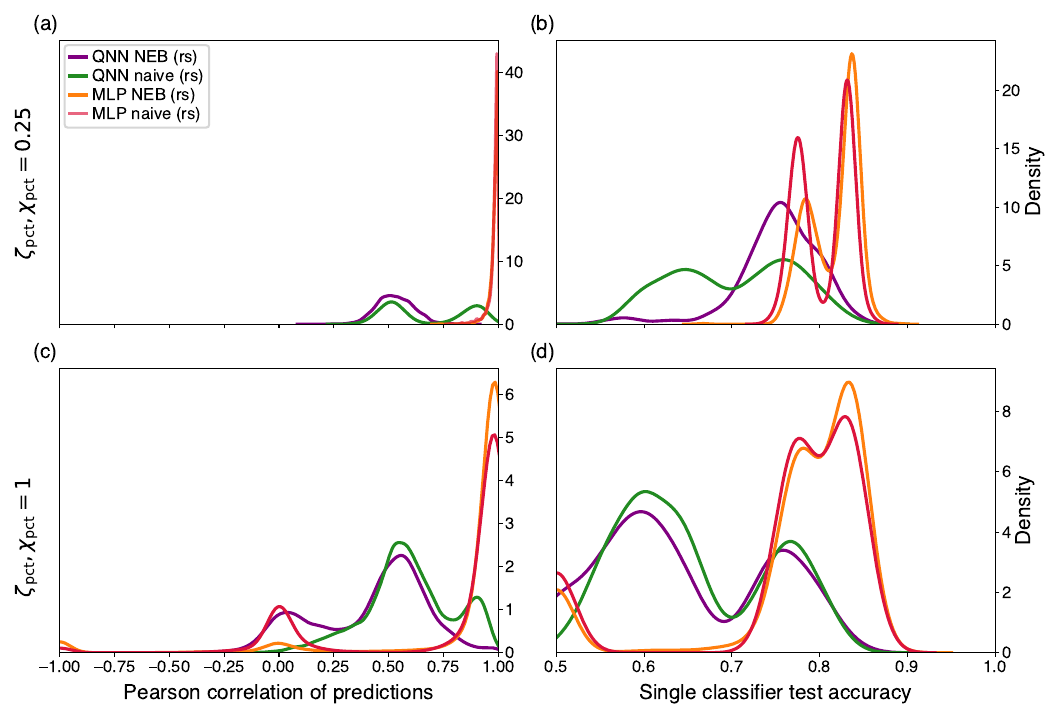}
    \caption{Distributions of the Pearson correlation coefficient between prediction vectors for all intra-ensemble pairs (\textbf{(a)},~\textbf{(c)}), and of the single-classifier test accuracy (\textbf{(b)},~\textbf{(d)}), filtered to ensembles in the 25th (\textbf{(a)},~\textbf{(b)}) and 100th (\textbf{(c)},~\textbf{(d)}) percentile of the prediction-correlation metric~$\zeta$, computed prior to optimization, or the \ac{mlp} validation cost filter $\chi$, computed after the optimization, respectively. Results are shown for all combinations of network type (\ac{qnn}, \ac{mlp}) and ensembling approach (\ac{neb}, naive). Here, the \acf{rs} \ac{neb} configuration was applied.}
    \label{fig:single_combined}
\end{figure}

\medskip \textbf{Independence of single classifiers}
Each pivot point corresponds to a parameter configuration and thus defines a baseline classifier. Figures~\ref{fig:single_combined}~(a, c) show the distribution of pairwise prediction-correlation coefficients across all pivots for one \ac{neb} optimization \textit{after} the optimization procedures, which is thus inherently different from the prediction-correlation metric $\zeta$, computed \textit{before} training. A positive correlation after training is found in almost all settings, which is expected since multiple baseline classifiers must agree in their predictions if they classify this state correctly. Nevertheless, the inter-prediction-correlation is substantially higher for both \ac{mlp} variants than for their \ac{qnn} counterparts, representing one of the key structural differences between the two network types. How far this observation generalizes beyond the present setting, and what theoretical mechanism underlies it, remain open questions that warrant further investigation.

Comparing the \ac{neb} and naive approaches in the quantum case, the \ac{qnn} \ac{neb} classifiers exhibit significantly lower pairwise correlation ($p<0.001$ for $\zeta_{\mathrm{pct}} \in \{0.25, 1.0\}$), which is beneficial for ensemble learning as described in Section~\ref{sec:ensemble-learning}. Further research is needed to establish whether this independence is an inherent property of ravines.

Statistical significance throughout this work is assessed using the nonparametric one-sided Mann-Whitney U test at a significance level of $0.05$. This test compares two arbitrary distributions and evaluates whether one is statistically smaller or greater than the other.

Note that the peak of near-zero correlation can be attributed to classifiers that do not surpass the baseline accuracy of $50\%$.

\medskip \textbf{Accuracy of single classifiers}
Turning to the second ingredient for strong ensemble performance, the accuracy of the individual baseline classifiers, the classical networks dominate as shown in Figures~\ref{fig:single_combined} (b,d). For low values of the validation cost filter metric ($\chi_\text{pct} = 0.25$), the \ac{mlp} classifiers exhibit two sharp peaks at high accuracy, outperforming the vast majority of \ac{qnn} baseline classifiers with $\zeta_\text{pct} = 0.25$, whose accuracy distribution is broad and diffuse. Nevertheless, the \ac{qnn} \ac{neb} classifiers achieve stronger individual performance than those from the naive approach ($p<0.001$).

Including all circuits and initializations ($\zeta_\text{pct}, \chi_\text{pct} = 1$), the picture becomes more ambiguous, with the notable exception of the naive \acp{qnn}: all other approaches include a non-negligible number of classifiers at the $50\%$ baseline accuracy. Furthermore, the dominance of the \ac{mlp} networks is less pronounced, as reflected in a greater overlap of the distributions. This is also a consequence of the \ac{mlp} filtering, which excludes poorly performing \ac{mlp} configurations from the $\chi_\text{pct} = 0.25$ subset.

Taken together, both filter metrics serve as strong indicators of base classifier performance, with the prediction-correlation metric $\zeta$ being particularly noteworthy given that it is computed \textit{prior to any} optimization, suggesting its utility as a general-purpose \ac{vqa} performance indicator beyond this study.

Interestingly, the correlation filter $\zeta$ appears to benefit the \ac{qnn} \ac{neb} approach even more strongly than the naive \ac{qnn} approach.
While both approaches seem to be optimized by taking this metric into account, the \ac{neb} protocol has the advantage that neighboring pivots collectively drive each other deeper into the ravine, requiring less iterations to build strong models than the naive protocol which trains each \ac{qnn} independently. A possible explanation is that $\zeta$ helps maximize performance for both approaches, but only the \ac{neb} approach leverages the valuable information of the initialized local minima to find further minima, which highlights the inherent advantage of our proposed protocol. 

\begin{figure}
    \centering
    \includegraphics[width=1\linewidth]{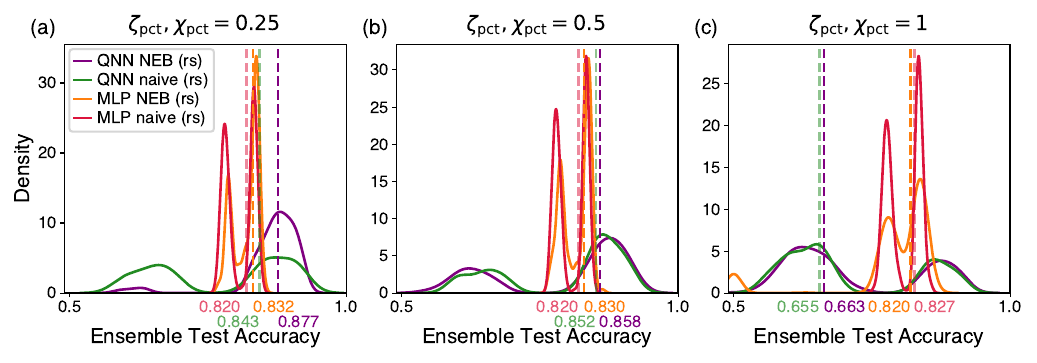}
    \caption{Ensemble test accuracy distributions for both network types (\ac{qnn} and \ac{mlp}) and ensembling approaches (\ac{neb} and naive), under three filtering thresholds: Subfigures~\textbf{(a)}, \textbf{(b)}, and~\textbf{(c)} correspond to the $25$th, $50$th, and $100$th percentile of the prediction-correlation metric $\zeta$, computed prior to optimization, or the \ac{mlp} validation cost filter $\chi$, computed after the optimization, depending on the network type. Here, the \acf{rs} \ac{neb} configuration was applied. Dashed vertical lines indicate the median of each distribution.}
    \label{fig:ensemble_combined}
\end{figure}

\medskip \textbf{Accuracy of the ensembles}
Combining the individual strength and independence of classifiers, the ensemble performance will be examined in the following. Figure~\ref{fig:ensemble_combined} shows the ensemble test accuracy distributions for all combinations of network types and optimization procedures across different threshold values $\zeta_{\mathrm{pct}}$ and $\chi_{\mathrm{pct}}$.

Most notably, the \ac{qnn} \ac{neb} approach outperforms all other classical and quantum methods for $\zeta_{\mathrm{pct}}, \chi_{\mathrm{pct}} \in \{0.25, 0.5\}$, thus providing superior ensemble performance while lowering resource demands compared to the naive \ac{qnn} by $\approx 27$\% as shown in Section~\ref{sec:absolute-complexity}. This demonstrates the high potential of the comparatively independent \ac{qnn} baseline classifiers and underlines the efficiency of the \ac{neb} approach in exploiting this independence. The previously stated performance differences are statistically significant ($p<0.001$) for all stated pairwise comparisons except for \ac{qnn} \ac{neb} and \ac{qnn} naive at $\zeta_{\mathrm{pct}} = 0.5$, where the two methods yield comparable results ($p=0.441$).

For $\zeta_{\mathrm{pct}}, \chi_{\mathrm{pct}} = 1$, both \ac{qnn} approaches are significantly outperformed by their classical \ac{mlp} counterparts ($p<0.001$), which can be attributed to the considerably larger performance spread of \acp{qnn} across different circuit designs and initializations. While this sensitivity to architectural and initialization choices can be a disadvantage when only a single circuit and initialization are available, it simultaneously presents an opportunity: the provided prediction-correlation metric allows for an informed selection of suitable circuits and initializations at barely any additional computational overhead.

Comparing both \ac{qnn} approaches directly, statistically significant performance differences are observed only for $\zeta_{\mathrm{pct}} = 0.25$, where the \ac{qnn} \ac{neb} approach surpasses naive ensembling ($p<0.001$). This is consistent with the earlier finding that \ac{neb} baseline classifiers exhibit a higher degree of independence in this regime compared to their naively trained counterparts, which in turn translates into a more diverse and therefore stronger ensemble. The single classifier accuracies are also higher for the \ac{qnn} \ac{neb} approach than for its naive counterpart. Nevertheless, this advantage diminishes for higher values of $\zeta_{\mathrm{pct}}$, suggesting that the \ac{qnn} \ac{neb} is particularly effective when base classifiers are drawn from well-initialized, low-$\zeta$ regions of the parameter space, where the \ac{neb} path exploration yields the most diverse set of solutions.

Furthermore, the data suggest an underlying bimodal structure in the performance distributions of both \acp{qnn} and \acp{mlp}. This is an artifact of the limited number of data sets studied and is expected to dissolve into a unimodal distribution in the limit of a large number of data sets and circuits. Additionally, the presumption that the dataset with CE values of $\{0.15, 0.35\}$ defines a harder learning problem compared to CE values of $\{0.05, 0.35\}$ is confirmed by the results.

In terms of absolute performance, all configurations except the weakest \ac{mlp} \ac{neb} achieve ensemble test accuracies well above the random baseline of $0.5$, demonstrating that the learning problem is consistently and reliably solvable across the methods studied. The highest observed median ensemble test accuracy of $\approx88\%$ further suggests that the task difficulty is well-calibrated: while the problem is tractable for the strongest classifiers, none of the methods reaches a performance ceiling, leaving room for further improvement and meaningful differentiation between approaches.

\subsection{Scaling analysis}
While the previous subsections provided evidence for the existence of ravines and their potential to be leveraged for stronger ensemble performance, this subsection analyses how these properties scale to deeper and wider circuits, thereby providing an indication of how the approach may apply in practical settings with large-scale, noise-resilient hardware. To gather the results presented below, only a subset of circuits and data shuffle seeds was used due to limited computational resources. Details on the experimental configuration are provided in Section~\ref{sec:experimental-details-scaling}.

\begin{figure}
    \centering
    \includegraphics[width=1\linewidth]{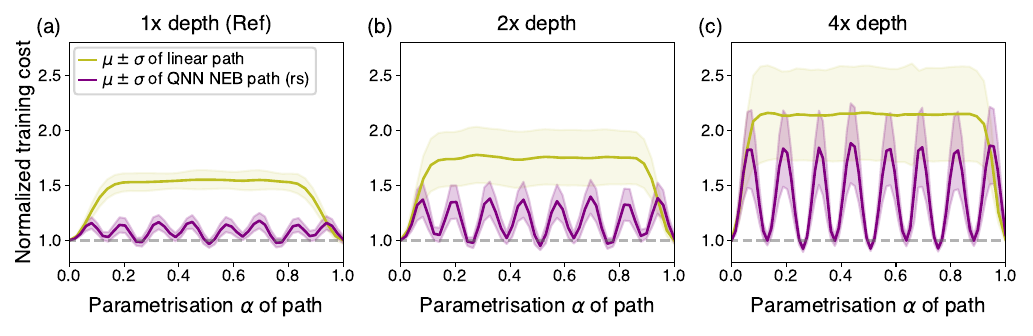}
    \caption{Training cost normalized to its value at $\alpha=0$, evaluated along both a linear interpolation and an optimized \ac{qnn} \ac{neb} path, for depth scaling factors of $1$ (reference), $2$, and $4$. Here, the \acf{rs} \ac{neb} configuration was applied.
    The solid lines represent the average across configurations, while the shaded areas indicate the corresponding $\pm 1\sigma$ band.}
    \label{fig:depth_average_path_combined}
\end{figure}

\medskip
\textbf{Depth scaling}
First, the scaling behavior with circuit depth is considered. Figure~\ref{fig:depth_average_path_combined} shows the linear and \ac{neb}-optimized paths for circuit depths scaled by factors of 1 (reference), 2, and 4. Note that the results shown, unlike those in the previous plots of this kind, originate from \ac{neb} runs of the \acf{rs} configuration, which gives rise to the characteristic spikes at the interpolation points between pivots.

\begin{figure}
    \centering
    \includegraphics[width=1\linewidth]{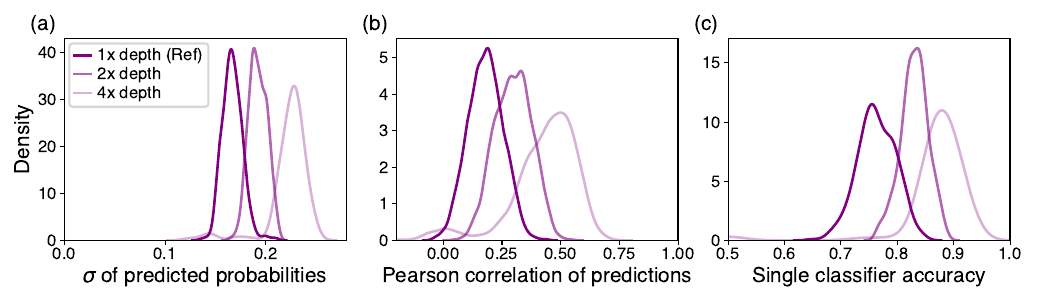}
    \caption{Distributions of single-classifier properties for the \ac{qnn} \ac{neb} ensemble across circuit depth scaling factors of $1$ (reference), $2$, and $4$: \textbf{(a)} the standard deviation of the predicted class probabilities across samples for a single classifier $\sigma$, \textbf{(b)} the Pearson correlation coefficient $R$ between prediction vectors for all intra-ensemble pairs, and \textbf{(c)} the single-classifier test accuracy. Each distribution is computed over two circuits and one data shuffle seed as described in section~\ref{sec:experimental-details-scaling}.}
    \label{fig:depth_single_combined}
\end{figure}

While these \ac{neb} runs cannot conclusively confirm the existence of ravines, since the path is sampled only at a small number of pivots, the resulting pivot configurations nonetheless exhibit the defining properties of ravines: low-cost local minima situated between the initialization points. Furthermore, a pronounced increase in the relative plateau height of the linearly connecting path is observed with increasing depth of the circuits. This can be attributed to the fact that deeper circuits admit local minima with considerably lower cost values, while randomly sampled parameter configurations yield approximately the same cost values regardless of circuit depth. This underlines that circuits with a larger number of parameters allow for a better approximation of the target classification function, and is further reflected in a systematic shift of the statistical properties of the individual pivot classifiers, as shown in Figure~\ref{fig:depth_single_combined}. As the spread of predicted probabilities across different samples increases with circuit depth (Fig.~\ref{fig:depth_single_combined}(a)), the classifiers exhibit higher decision confidence, which in turn enables them to reach lower training cost values (Fig.~\ref{fig:depth_single_combined}(c)). This improvement in predictive quality, however, comes at the cost of increased correlation between the predictions of different classifiers, which is precisely the trend observed in Fig.~\ref{fig:depth_single_combined}(b).

\begin{figure}
    \centering
    \includegraphics[width=0.45\linewidth]{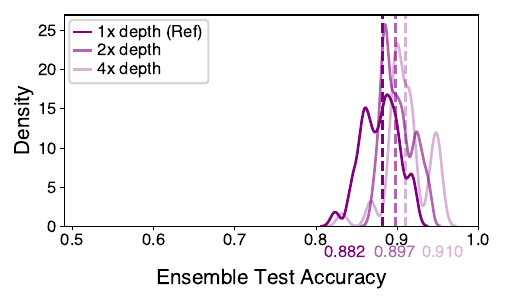}
    \caption{Ensemble test accuracy distributions of \ac{qnn} \ac{neb} ensembling for depth scaling factors of $1$ (reference), $2$, and $4$. Dashed vertical lines indicate the median of each distribution.}
    \label{fig:depth_ensemble}
\end{figure}

Combining the individual classifiers into ensembles confirms the overall trend of improving performance with increasing circuit depth: Figure~\ref{fig:depth_ensemble} shows that the ensemble test accuracy rises steadily with circuit depth, reaching up to $91\%$ for the deepest configuration studied. However, this performance gain comes at a commensurate increase in computational cost, as the complexity of a single gradient evaluation scales linearly with the number of parameters meaning that the 2x and 4x depth configurations require roughly a factor of 2 or 4 more computational resources, respectively, for only modest accuracy improvements. This highlights an important practical consideration: slightly weaker baseline classifiers operating at shallower depth can yield comparable ensemble performance provided their predictions are sufficiently independent, effectively achieving similar results at a fraction of the computational cost. Beyond this trade-off, the results demonstrate that the \ac{qnn} \ac{neb} approach scales favorably with circuit depth and retains strong ensemble performance across all circuit regimes studied.

\medskip
\textbf{Qubit scaling}
Lastly, the scaling of the \ac{neb} method with circuit width is analyzed, as this constitutes the crucial scaling factor for large-scale \ac{qml} applications. To conduct this comparison, different datasets are required for different qubit counts, rendering a direct comparison of absolute performance values impractical. Nevertheless, the existence of ravines and the amount of computational resources required for convergence can still be analyzed to evaluate the scaling prospects of the \ac{neb} approach.

\begin{figure}
    \centering
    \includegraphics[width=\linewidth]{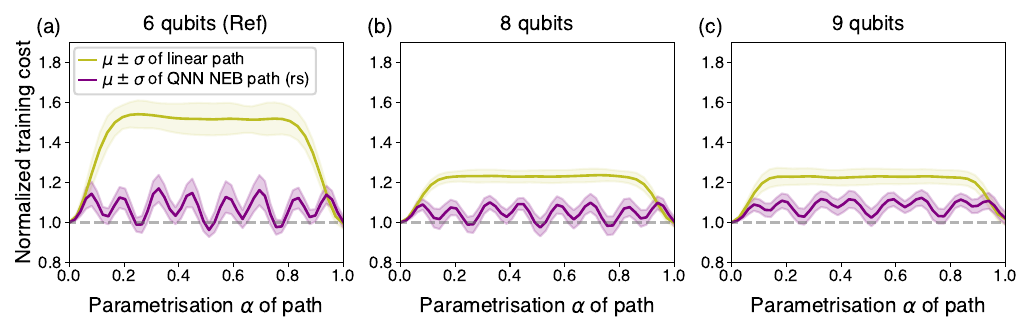}
    \caption{Training cost normalized to its value at $\alpha=0$, evaluated along both a linear interpolation and an optimized \ac{qnn} \ac{neb} path, for circuit widths of $6$ qubits (reference), $8$ qubits, and $9$ qubits. Here, the \acf{rs} \ac{neb} configuration was applied.
    The solid lines represent the average across configurations, while the shaded areas indicate the corresponding $\pm 1\sigma$ band.}
    \label{fig:width_average_path_combined}
\end{figure}

Figure~\ref{fig:width_average_path_combined} shows the linear and \ac{neb}-optimized paths for circuits of 6 (reference), 8, and 9 qubits. Two observations stand out: first, the \ac{neb} pivots significantly reduce the cost values along the linear connection, presumably in the process of converging towards a ravine-like structure; and second, the plateaus along the linear connection are less pronounced than in the 6-qubit reference case. The latter can be explained by the fact that the local minima used as initialization points exhibit generally higher cost values for wider circuits, while randomly sampled points in parameter space yield approximately the same cost values as in the reference case.

The fact that the pivot points have not yet reached the cost level of their respective initialization points suggests that the \ac{neb} procedure requires additional computational resources for wider circuits. This raises the question of how much extra resource expenditure is needed to compensate for the increased number of qubits. To address this, the number of optimization steps in the \ac{neb} procedure was increased for the 8- and 9-qubit convergence experiments, allowing for a more accurate analysis of the convergence behavior. Further details on the experimental setup are provided in Section~\ref{sec:experimental-details-scaling}.

\begin{figure}
    \centering
    \includegraphics[width=.5\linewidth]{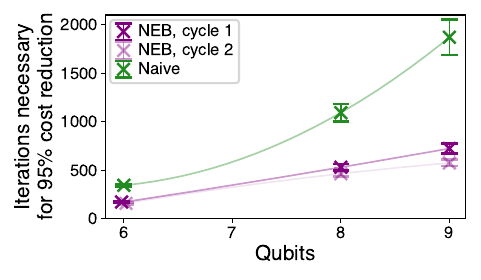}
    \caption{Number of iterations required to achieve a $95\%$ reduction in \ac{qnn} training cost relative to the convergence limit estimated by an exponential fit. Details on the calculation are provided in Section~\ref{sec:experimental-details-scaling}. \ac{neb} cycle 1 and 2 refer to the set of pivots which were added in cycle 1 and 2 in the adapted \ac{neb} procedure, respectively. A quadratic interpolation is added for visualization purposes.}
    \label{fig:width_resource_scaling}
\end{figure}

\begin{figure}
    \centering
    \includegraphics[width=0.5\linewidth]{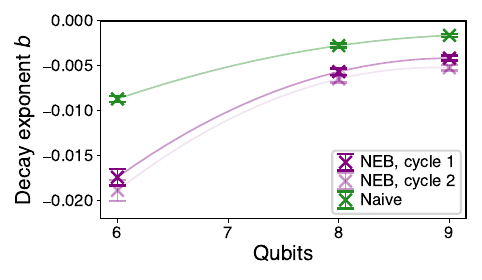}
    \caption{Decay exponent $b$ of the exponential fit $f(x) = ae^{bx} + c$ to the average \ac{qnn} training cost decay, as described in Section~\ref{sec:experimental-details-scaling}. \ac{neb} cycle 1 and 2 refer to the set of pivots which were added in cycle 1 and 2 in the adapted \ac{neb} procedure, respectively. A quadratic interpolation is added for visualization purposes.}
    \label{fig:width_exponent_scaling}
\end{figure}

The metric $\mathrm{it}^{\star}$ introduced in Section \ref{sec:experimental-details-scaling} is a measure for iterations necessary for convergence to a convergence point parameterized by $\delta$. For $\delta=0.05$, the resulting values of $\mathrm{it}^{\star}$, shown in Figure~\ref{fig:width_resource_scaling}, exhibit a substantial increase with the number of qubits for both approaches. Nevertheless, the resources required for convergence are consistently roughly twice as large for the naive approach as for the \ac{neb} approach, a trend that holds across all qubit counts studied. Furthermore, the resources required for convergence in the second \ac{neb} cycle are slightly below those of the first cycle, suggesting that previously explored landscape structure is effectively reused. The values for the exponential decay parameter $b$ are shown in Figure~\ref{fig:width_exponent_scaling}.

Taken together, these results suggest that the \ac{neb} method scales at least as favorably as the naive approach. In the specific configurations studied, the \ac{neb} approach even exhibited faster convergence, which may be attributed to the fact that it combines gradient information from multiple pivots simultaneously, and thereby effectively leverages the geometrical structure of \acp{qcl}.

\section{Limitations}
\label{sec:limitations}
In this study, we conducted and analyzed several thousand optimization procedures using both \ac{neb} and naive ensembling across multiple datasets, data splits, circuit designs, including various depths and widths, and initializations to improve generalizability. Numerical studies are, however, inherently limited to the data they generate: here, circuits of up to $9$ qubits and $288$ parameters. While modest on the scale of long-term applications, this scope was constrained by the computational cost of producing a statistically-meaningful amount of optimization runs. The simulations further assumed error-free gradient evaluation. Although current hardware does not satisfy this requirement, approaches such as the one presented here are intended for deployment once quantum hardware becomes fast enough for meaningful machine learning applications.

A further limitation of the proposed \ac{aneb} ensembling framework is its dependence on the presence of ravine structures in the cost landscape. Although our observations of ravines are consistent with prior work (\cite{kathleene.hamilton2021}), their existence has not been established for arbitrary circuit configurations. In cases where local minima are unconnected by such structures, the \ac{neb} procedure will fail to identify low-cost connecting paths, potentially yielding ensemble predictions inferior to the naive baseline.

Finally, the filtering metric, which substantially improves \ac{qnn} ensemble performance, may pose practical difficulties in settings where the circuit architecture is fixed or resampling is not feasible.

\section{Outlook}
\label{sec:outlook}
Promising directions for future work include scaling this study to circuits of greater depth and width to verify that ravine structures persist beyond the parameter regimes accessible here. Complementing these empirical investigations with theoretical analysis would be equally valuable: a rigorous treatment could establish conditions for the existence of ravines and explain the notably-high independence observed among \acp{qnn} in general and within the \ac{neb} ensembling framework in particular. Such theoretical grounding would deepen our understanding of fundamental differences between quantum and classical machine learning.

Another natural extension is a systematic comparison of \acp{qcl} with their classical counterparts regarding properties beyond ravines. The structure of local minima and their relationship to architecture choices has been studied extensively in classical settings (\cite{keskar2017, draxler2019}). Analogous investigations in the quantum regime remain largely open. While ravines already enable the tailored optimization procedures demonstrated here, a broader characterization of \ac{qcl} properties could inform the design of additional improvements to \ac{qnn} training.

Finally, given that the prediction-correlation metric $\zeta$ is such a resource-light metric while being such a strong indicator of \ac{vqa} performance, further research on the underlying mechanism of this phenomenon, as well as on its generalization to different \ac{qml} models, are important future directions.

\section{Conclusion} \label{sec:conclusion}
This study investigated the structure of \acp{qcl} across a broad range of settings with respect to low-cost connections between local minima, commonly referred to as ravines (\cite{kathleene.hamilton2021}). The \ac{neb} algorithm identified ravines in the vast majority of configurations, extending previous findings to quantum data and to circuits with up to $9$ qubits and $288$ variational parameters.

Individual pivot points along the \ac{neb} low-cost paths were shown to exhibit greater prediction independence than both quantum and classical baselines.
The \ac{aneb} ensembling procedure leverages this independence to construct a strong ensemble classifier, outperforming both quantum and classical alternatives when combined with a lightweight circuit initialization and filtering procedure prior to optimization. The introduced filtering metric itself proved to be a strong indicator of \ac{vqa} performance more broadly, suggesting that its applicability extends well beyond the scope of this study.
Our complexity analysis further shows that \ac{aneb} achieves this while requiring substantially fewer resources than the naive ensemble approach. Finally, the depth and qubit scaling analysis provided promising evidence that ravines persist in larger circuits, and that the \ac{aneb} convergence scales more favorably in terms of resource requirements than naive \ac{qnn} ensembling.

Taken together, these results demonstrate that \ac{qcl}-structure-informed algorithms can improve ensemble performance while reducing computational cost. The exploitation of landscape geometry is therefore a compelling direction for future research, with applications ranging from tailored optimization procedures to warm-starting strategies.

\backmatter



\bmhead{Data Availability}
The data used in this study are available upon reasonable request. Equivalent datasets can be generated following the procedure of \cite{schatzki2021} with the code available at \url{https://github.com/LSchatzki/NTangled_Datasets}.

\bmhead{Funding}
This work was supported by the Dieter Schwarz Foundation through the Fraunhofer Heilbronn Research and Innovation Centers HNFIZ.

\bmhead{Competing Interests}
The authors declare no competing interests.



\begin{appendices}

\section{\Acl{neb} algorithm}
\label{sec:appendix-alg-neb}
We present the pseudocode of the \acf{neb} algorithm introduced by \cite{jonsson1998} and \cite{henkelman2000}, introduced in Section~\ref{sec:neb} as Algorithm~\ref{alg:neb}. It appears as a subroutine of the \acf{aneb} algorithm (Algorithm~\ref{alg:adapted-neb}).

\begin{algorithm}
\caption{\ac{neb} (\cite{jonsson1998, henkelman2000})}\label{alg:neb}
\begin{algorithmic}[1]
\Require{Initial path $\mathbf{p} = [\mathbf{p}_0, \mathbf{p}_{1}, \mathbf{p}_{2}, \dots, \mathbf{p}_{N+1}]$ with $N+2$ pivots}
\Function{NEB}{$\mathbf{p}, \eta, T, k)$}
    \For{$t = 1, \dots ,T$}
        \For{$i = 1, \dots, N$}
            \State Compute combined force $\mathbf{F}{(\mathbf{p}_i}, k) =  \mathbf{F}^{\mathrm{S}} (\mathbf{p}_{i}, k) - \mathbf{\nabla} \mathcal{C}(\mathbf{p}_{i})\big |_{\perp}$
            \State Store pivot $\mathbf{p}_{i}^{(t)} = \mathbf{p}_{i}^{(t-1)} +  \eta \, \mathbf{F}{(\mathbf{p}_i}, k)$
        \EndFor
    \EndFor
    \State \Return $\mathbf{p}^{(T)}$
\EndFunction
\end{algorithmic}
\end{algorithm}

\section{Experimental details}
Here we detail the experimental configuration necessary to reproduce the presented results. Table~\ref{tab:ds-parameters} and \ref{tab:4q-ds-parameters} provides details on the data set generation hyperparameters following the procedure from \cite{schatzki2021} for 6 and 8 qubits, respectively.
Table~\ref{tab:neb-parameters} includes the \acf{aneb} hyperparameter choices which are motivated in section \ref{sec:training-protocols} and Table~\ref{tab:scheduler-parameters} specifies the step scheduler configuration used in the \ac{aneb} workflow introduced in Section~\ref{sec:neb}.

\begin{table}[h!]
    \centering
    \begin{tabular}{c||c|c}
          & Data set 1 & Data set 2 \\
         \hline
         \hline
         $(\mathrm{CE}_1, \mathrm{CE}_2)$ & $(5,35)$ & $(15, 35)$ \\
         \hline
         Generating architecture & \multicolumn{2}{c}{Hardware efficient}\\
         \hline
         Generating circuit depth & 4 & 3 \\
         \hline
         Generating input states & \multicolumn{2}{c}{Product states}
    \end{tabular}
    \caption{Details on the parameters used for 6 qubit data set generation.}
    \label{tab:ds-parameters}
\end{table}

\begin{table}[h!]
    \centering
    \begin{tabular}{c||c|c}
          & Data set 1 & Data set 2 \\
         \hline
         \hline
         $(\mathrm{CE}_1, \mathrm{CE}_2)$ & $(5,35)$ & $(15, 35)$ \\
         \hline
         Generating architecture & \multicolumn{2}{c}{Hardware efficient}\\
         \hline
         Generating circuit depth & \multicolumn{2}{c}{6} \\
         \hline
         Generating input states & \multicolumn{2}{c}{Product states}
    \end{tabular}
    \caption{Details on the parameters used for 8 qubit data set generation.}
    \label{tab:4q-ds-parameters}
\end{table}

\begin{table}[h!]
    \centering
    \begin{tabular}{c||c|c}
        & \makecell{resource-scarce \\ configuration} & \makecell{resource-abundant \\ configuration}\\
        \hline
        \hline
        Number of initialization points & 5 & 11 \\
        \hline
        Number of pivots added per cycle & 4 & 2 \\
        \hline
        Cycles & 2 & 10 \\
        \hline
        Number of interpolation points $M$ & \multicolumn{2}{c}{9} \\
        \hline
        Step size $\gamma$ & 0.2 & 0.1 \\
        \hline
        Spring constant $k$ & 1.0 & 0.0\\
        \hline 
        Method to add new pivots & linear & threshold \\
        \hline
        Deviation threshold & / & $1 \cdot 10^{-2}$  \\ 
    \end{tabular}
    \caption{Hyperparameter configurations for the adapted \ac{neb} algorithm. The resource-scarce configuration prioritizes computational efficiency, while the resource-abundant configuration prioritizes convergence to a low-cost path.}
    \label{tab:neb-parameters}
\end{table}

\begin{table}[h!]
    \centering
    \begin{tabular}{c||c|c}
        & \makecell{resource-scarce \\ scheduler} & \makecell{resource-abundant \\ scheduler}\\
        \hline
        \hline
        Initial steps & 300 & 1000 \\
        \hline
        Minimum number of steps & 100 & 200 \\
        \hline
        First decay rate & 0.7 & 0.6 \\
        \hline
        Standard decay rate & 0.8 & 0.8 \\
    \end{tabular}
    \caption{Step scheduler configurations for the two \ac{neb} settings. The first cycle uses more iterations in both cases to obtain an initial approximation of the cost ravine, which is subsequently refined.}
    \label{tab:scheduler-parameters}
\end{table}




\end{appendices}

\newpage
\bibliography{sn-bibliography}

\end{document}